\documentclass[reprint,aps,pra,two-column,showpacs]{revtex4-2}
\usepackage{bm,amsmath,amssymb,amsfonts}
\usepackage{color}
\usepackage{dcolumn}
\usepackage{graphicx}
\usepackage{float}
\usepackage{hyperref}
\hypersetup{colorlinks=true,citecolor=blue,urlcolor=blue,linkcolor=blue}
\usepackage{ulem}
\usepackage{times}
\usepackage{subfigure}
\usepackage{appendix}

\begin{document}
\setlength{\abovedisplayskip}{12pt} 
\setlength{\belowdisplayskip}{12pt}
\title{Engineering anisotropic Dicke model with dipole-dipole interaction for Rydberg atom arrays in cavity}
\author{Bao-Yun Dong\textsuperscript{1,2}}
\author{Yanhua Zhou\textsuperscript{4,5}}
\author{Wei Wang\textsuperscript{1,2}}
\thanks{corresponding author: weiwangphys@163.com}
\author{Tao Wang\textsuperscript{1,2,3}}
\thanks{corresponding author: tauwaang@cqu.edu.cn}
\affiliation{1\, Department of Physics, and Chongqing Key Laboratory for Strongly Coupled Physics, Chongqing University, Chongqing, 401331, China}
\affiliation{2\, Center of Modern Physics, Institute for Smart City of Chongqing University in Liyang, Liyang, 213300, China}
\affiliation{3\, Center of Quantum Materials and Devices, Chongqing University, Chongqing 401331, China}
\affiliation{4\, National Time Service Center, Chinese Academy of Sciences, Xi’an 710600, People’s Republic of China}
\affiliation{5\, School of Astronomy and Space Science, University of Chinese Academy of Sciences, Beijing 100049, China}

\begin{abstract}
The anisotropic Dicke model reveals the important role that counter-rotating wave terms play in the coupling between light and two level atoms. It is intriguing to generate the model to the strongly correlated many body case, where the competition between atomic interaction and light-atom coupling will induce exotic phenomenon. In this paper, we provide a periodically driving method for engineering an anisotropic Dicke model with strong dipole-dipole interaction between atoms based on Rydberg atom arrays in cavity. By modulating the pumping laser, the ratio of counter-rotating wave terms to rotating ones could be tuned from zero to infinity. 
As an illustrative example, the superiority of this tunability in the adiabatic state preparation is revealed. 
Our engineered model provides an extensive foundation for subsequet research of quantum simulation, many-body physics and even information processing.
\end{abstract}
\maketitle

\section{INTRODUCTION}
As one of correlated many-body systems, the light-matter interaction problem has attracted considerable attention and has been actively investigated over many decades\cite{Kockum, Gutzler}.
It is desired to capture the underlying physics of such issues. 
The Dicke model was proposed in 1973 by K.Hepp and E.H.Liep\cite{Hepp}, and describes a large number of two-level atoms coupled to a single mode of a quantized radiation field.
When the coupling strength is much smaller than the mode frequency, there is a simplified scenario.
Neglecting the counter-rotating wave(CRW) terms, the reduced Tavis-Cummings model \cite{Agarwal, Fink, Restrepo, Feng, Yang} exhibits excitation number conservation.
On the other hand, if the rotating wave approximation (RWA)\cite{Kazuyuki} fails and the system reaches the strong coupling regime\cite{Garbe, Liberti, Forn, Jaako}, the full quantum Dicke model\cite{Kirton,Baumann,Chiacchio,Das,Garraway,Hassan} is necessary.
The investigation of CRW terms illustrates that many interesting quantum effects \cite{Niemczyk,Cao,Garziano,Wang,Casanova,Wangxin, Plium} can be induced, such as non-classical states preparation\cite{Ashhab}, three-photon resonances\cite{Ma}, the deterioration of photon blockade effect\cite{LeB,Hwang}, and so on.

Recently, the generalized Dicke model, named the anisotropic Dicke model (ADM)\cite{adm1, adm2, adm3, adm4}, has sparked considerable research interest, whose impressive feature is the different coupling strengths for the rotating wave (RW) terms and CRW terms. 
Because of these intriguing properties, it has been employed to investigate a range of theoretical topics, including quantum phase transitions\cite{Dasp,Zhuxin}, critical behaviors\cite{Zhuxin} and Floquet heating\cite{Daspra}. For example, exciting research results indicate that the ADM exhibits not only a phase transition from the normal phase to the superradiant(SR) phase but also ergodic to nonergodic transitions\cite{Buijsman}.
It is noteworthy that the observation of those intriguing physical phenomena of such ADM largely relies on the tunability of CRW terms. 
Moreover, the high controllability of the RW and CRW terms coupling strengths provides support for observing the effects of the pure CRW terms.
However, how to engineer the tunability in real experiments is still an open question to date.

Quantum simulation utilizes controlled quantum systems to model complex physical phenomena that are intractable classically, aiming to explore quantum many-body problems\cite{Georgescu,Altman,Daley}. 
Among these simulators, Rydberg atoms offer a unique experimental platform due to their advantages in scalability, coherence time, and strong, tunable interactions\cite{ryd2,ryd3,ryd4,ryd1}. 
Placing the atomic ensemble within an optical cavity allows for the long-range interaction between atoms and the cavity, making it possible to create multi-partite entanglement and observe collective phenomenon\cite{cav1,cav2,cav4,cav5,cav6}.
Recent advances in the programmable manipulation of isolated systems have enabled detailed studies of equilibrium and nonequilibrium states under strong matter-matter interactions\cite{qph,nond1,qcco}.
Remarkably, investigating quantum adiabatic evolution is a promising way to observe quantum many-body dynamics\cite{nond2} and solve combinatorial optimization problems\cite{QOW, bomb}.
It is then a next step to approach the interplay between strong light-atom and atom-atom interactions\cite{cav7, isingm}, which represents significant advancement in simulating and understanding relative many-body problems of quantum matter and light.

In this paper, we propose to realize an anisotropic Dicke model and Floquet engineering its CRW in the context of Rydberg array simulations. 
Specifically, by designing an appropriate two-photon process of three-level atoms, we derive an effective two-level Hamiltonian incorporated into strong atom-interaction ADM.
The light-atom coupling is capable of reaching both the strong coupling and ultra-strong coupling regimes with the high-finesse optical cavity. 
We further show the tunability of CRW can be engineered by modulating the pump light. 
As an application of this controllability, we demostrate the dependence of the distribution of eigenenergies on it could facilitate adiabatic SR and superradiant solid state(SRS) state preparation. 
Here the SRS phase is novelly characterized by the coexistence of solid and superradiance order, and has been predicted due to the competition between the nearest-neighbor Rydberg blockade and the long-range exchange interactions mediated by the photon\cite{srs1,srs2,srs3}.
It provides the necessary conditions for the study of related phases.

The remainder of this paper is organized as follows. 
Sections \ref{Floquet} first introduces the theoretical model and our driven scheme. 
We then derive the effective anisotropic Dicke Hamiltonian and discuss its tunability. 
Demonstrated with SR and SRS phases respectively, 
Sec. \ref{Adiabatic} is dedicated to analyzing the impact of controllable strong CRW terms on their adiabatic preparation. 
Lastly, in Sec. \ref{outlook}, we conclude with a discussion.

\section{Floquet engineering effective Hamiltonian}
\label{Floquet}

\begin{figure}[t]
    \centering
    \includegraphics[width=1.10\linewidth]{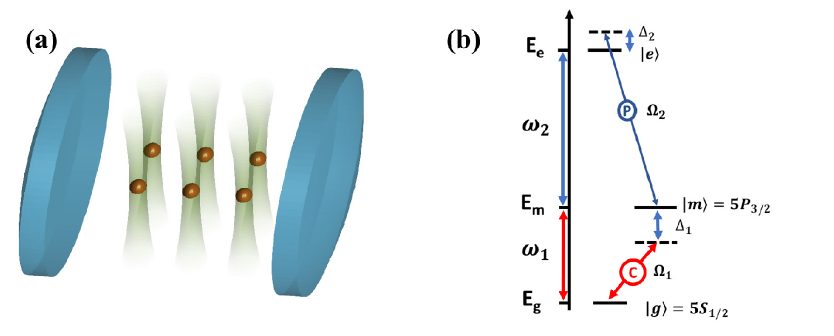}
    \caption{(a) Schematic diagram of Rydberg atoms trapped by square lattice optical tweezers, which interact with an ultrahigh finesse optical cavity. 
    Each site of arrays is occupied by a single atom and located at the antinode of the cavity mode. 
    (b) Illustration of atomic energy levels and the associated two-photon pathway. 
    The cavity mode excites Rubidium atomic ground states $5S_{1/2}$ to intermediate states $5P_{3/2}$.
    $5P_{3/2}$ are coupled to Rydberg state $|e \rangle$ by a beam of classic laser.
    After introducing driven microwave fields and removing intermediate states, the two-photon processes assisted with Flquet photon give rise to the modulated coupling between photons and two-level atoms.}
    \label{fig1}
\end{figure}

We discuss cold Rydberg atoms trapped by optical tweezers with geometry that forms a one-dimensional chain or two-dimensional square lattice. 
Placed inside an ultra-finesse optical cavity, the atoms are further assumed to couple to the cavity photon and a beam of classical laser, which is schematically depicted in Fig.\ref{fig1}(a).
Considering each atom at the antinode of the cavity mode, the atomic ground states $| g \rangle$ and the intermediate states $| m \rangle$ are coupled by a single quantized cavity mode with the strength $\Omega_{1}$. 
The detuning is $\Delta_1=\omega_{1}-\omega_{c}$, where $\omega_{c}$ is the frequency of the cavity mode, and $\omega_{1}$ is the frequency difference between $| g \rangle$ and $| m \rangle$.
A classical laser field of frequency $\omega_{p}$  non-resonantly drives the transition of $| m \rangle\to | e \rangle$ with the Rabi frequency $\Omega_{2}$. Denoting $\omega_2$ as the frequency difference between $| m \rangle$ and $| e \rangle$, the detuning is $\Delta_2=\omega_2-\omega_p$.
The above two-photon process is shown in Fig.\ref{fig1}(b), where the total detuning is $\Delta=\Delta_1+\Delta_2$.
Omitting atomic external motions and setting $\hbar=1$, one can describe the dynamics by the following Hamiltonian
\begin{align}
	H_{E} = H_{D} + H_{I} \label{E1}
\end{align}
The diagonal term $H_{D}$, and the interaction terms $H_{I}$ are correspondingly given as
\begin{align} 
    H_{D}=&\omega_c\hat{a}^{\dagger}\hat{a}+\omega_1\sum_{j=1}^{N} | m_j \rangle \langle m_j |+(\omega_1+\omega_2)\sum_{j=1}^{N} |e_j \rangle \langle e_j | \nonumber\\
    +& V\sum_{<i,j>}|e_i\rangle \langle e_i|e_j \rangle \langle e_j | \label{E2}\\
    H_{I} =& \frac{\Omega_{1}}{\sqrt{N}}(\hat{a}^{\dagger}+\hat{a})\sum_{j=1}^{N}( | m_j \rangle \langle g_j |+| g_j \rangle \langle m_j | ) \nonumber\\
    +& \frac{\Omega_{2} \mathrm{cos(\omega_pt)}}{\sqrt{N}}\sum_{j=1}^{N}(| m_j \rangle \langle e_j |+| e_j \rangle \langle m_j |) \label{E3}
\end{align}
Here $a^{\dagger}$($a$) is the creation (annihilation) operator of photons in the cavity.
$V$ is the strength of the nearest-neighbor dipole-dipole interaction of the Rydberg state.
$N$ is the total number of atoms trapped by optical tweezers.  
The maximum occupation number of atoms in each site should be one because of the Rydberg blockade.

To deepen our understanding of the CRW terms, hereafter we floquet engineering the Hamiltonian.
To be specific, we apply a beam of microwave field to the system, modulated by angular frequency as $\omega_{l}(t) = \bar{\omega} + \omega_a \sin(\omega_s t)$, where $\bar{\omega}$ is the frequency of microwave without driving and $\omega_a$ is the driving amplitude\cite{floq}.
So there is an equivalent potential $A\cos{\omega_{s}t}$ on the Rydberg states with $A$ the amplitude of potential. 
Then the Hamiltonian of the diagonal part is modified as

\begin{align} 
    H_{Dri}=&\omega_c\hat{a}^{\dagger}\hat{a}+\omega_1\sum_{j=1}^{N} | m_j \rangle \langle m_j |
     + V\sum_{<i,j>}|e_i\rangle \langle e_i|e_j \rangle \langle e_j |\nonumber\\
    +&(\omega_1+\omega_2+A\cos{\omega_s t})\sum_{j=1}^{N} |e_j \rangle \langle e_j | \label{EDri1}
\end{align}

In order to deal with the time-dependent terms in Eq.\eqref{E3}, we implement a unitary transformation by the operator $u=\rm exp\{\sum_{j=1}^{N}(-i|e_{j} \rangle \langle e_{j} | \omega_{P}t) \}$.
After employing the RWA to the interaction between the classical laser and atoms, the Hamiltonian becomes
\begin{align}
     H_{DR}=&\omega_c\hat{a}^{\dagger}\hat{a}+\omega_1\sum_{j=1}^{N} | m_j \rangle \langle m_j |
     + V\sum_{<i,j>}|e_i\rangle \langle e_i|e_j \rangle \langle e_j |\nonumber\\
     +&(\omega_1+\Delta_{2}+A\cos{\omega_{s}t})\sum_{j=1}^{N} |e_j \rangle \langle e_j | \label{E4}\\
     H_{IR} =& \frac{\Omega_{1}}{\sqrt{N}}(\hat{a}^{\dagger}+\hat{a})\sum_{j=1}^{N}( | m_j \rangle \langle g_j |+| g_j \rangle \langle m_j | ) \nonumber\\
    +& \frac{\Omega_{2}}{\sqrt{N}}\sum_{j=1}^{N}(| m_j \rangle \langle e_j | +| e_j \rangle \langle m_j | ) \label{E5}
\end{align}
In a typical Rydberg experiment, using $\rm ^{87}Rb$ atoms as an example, the intermediate states, $5P_{3/2}$, have a long lifetime(denoting as $\tau$).
Detuning the cavity mode such that $\Delta \approx 0$ and $\Delta_{1} \gg \tau^{-1}$, then the intermediate states can be regarded as unpopulated and can be eliminated to reduce the effective Hamiltonian. 
We follow the standard Schrieffer-Wolff transformation\cite{swt}.
Since the periodic driving contributes zero to the shift of energy levels on average over a long period of time, we first construct an antihermitian  generator $S$ based on interaction Hamiltonian Eq.\eqref{E5} for each site $j$ as $ S^{j} = \left (\frac{\Omega_{1}a|m_{j}\rangle\langle g_{j}|}{\Delta_{1}} + \frac{\Omega_{1}a^{\dagger}|m_{j}\rangle\langle g_{j}|}{\omega_{1}+\omega_{C}} -\frac{\Omega_{2}|m_{j}\rangle\langle e_{j}|}{\Delta_{2}} \right )-h.c.$
With the second order truncation, we next reach the subspace effective Hamiltonian $H_{eff}^{j}=e^{S}(H_{DR}^{j}+H_{IR}^{j})e^{-S}=H_{DR}^{j}+\frac{1}{2}[S^{j},H_{IR}^{j}]$, 
where the commutation relation can be straightforwardly formulated as
\begin{widetext}
\begin{align}
    \left[ S^{j}, H_{int}^{j} \right] &=|\Omega_{1}|^{2}|m_{j}\rangle\langle m_{j}|\left ( \frac{2a^{\dagger}a}{\omega_{1}+\omega_{C}}+\frac{a^{\dagger 2}}{\omega_{1}+\omega_{C}}+\frac{a^{\dagger 2}}{\Delta_{1}}+\frac{a^{2}}{\Delta_{1}}+\frac{a^{2}}{\omega_{1}+\omega_{C}}+\frac{2aa^{\dagger}}{\Delta_{1}} \right )-\frac{2|\Omega_{2}|^{2}}{\Delta_{2}} |m_{j}\rangle\langle m_{j}|\label{ES1}\\
    &-| \Omega_{1}|^{2} | g_{j} \rangle \langle g_{j} | \left ( \frac{2aa^{\dagger}}{\omega_{1}+\omega_{C}} + \frac{a^{2}}{\omega_{1}+\omega_{C}} + \frac{a^{2}}{\Delta_{1}} + \frac{a^{\dagger 2}}{\Delta_{1}} + \frac{a^{\dagger 2}}{\omega_{1}+\omega_{C}} + \frac{2a^{\dagger}a}{\Delta_{1}}\right )
    +\frac{2|\Omega_{2}|^{2}}{\Delta_{2}}|e_{j} \rangle \langle e_{j} | \label{ES3}\\
    &+\Omega_{1}^{*} \Omega_{2} | g_{j} \rangle \langle e_{j} | \left ( -\frac{a}{\omega_{1}+\omega_{C}} + \frac{a}{\Delta_{2}} - \frac{a^{\dagger}}{\Delta_{1}} + \frac{a^{\dagger}}{\Delta_{2}}\right )
    +\Omega_{2}^{*} \Omega_{1} | e_{j} \rangle \langle g_{j} |\left ( -\frac{a^{\dagger}}{\omega_{1}+\omega_{C}} + \frac{a^{\dagger}}{\Delta_{2}} - \frac{a}{\Delta_{1}} + \frac{a}{\Delta_{2}} \right)\label{ES5}
\end{align}
\end{widetext}

Now we proceed to discuss a simpler result in which one only cares about atomic dynamics for such atom-photon interaction systems.
In this case, we drop Eq.\eqref{ES1}-\eqref{ES3} because those second-order photon interactions only modify the energy levels of atomic states. 
For the leaving terms in Eq.\eqref{ES5}, it is clear that the intermediate state $m_{j}$ is decoupled from the ground state $g$ and the Rydberg state $e$.
Finally the effective Hamiltonian can be expressed as a two-level system,
\begin{align}
    H_{eff}^{j} =&\omega_{C}a^{\dagger}a + (\omega_{A}+A\cos{\omega_{s}t}) n_{j} \nonumber\\
    +&\left (\Omega_{e1} a\sigma^{(+)}_{j}+\Omega_{e2} a^{\dagger} \sigma^{(+)}_{j} + h.c.\right ) \label{ee1}
\end{align}
Note that $\omega_{1}$ and $\omega_{C}$ can be in the hundreds of $\rm THz$ level in the Rubidium atoms experiments, we neglect $\frac{1}{\omega_{1}+\omega_{C}}$. Then, we define the CRW and RW effective coupling strengths as $\Omega_{e1}=\frac{\Omega_{2}^{*}\Omega_{1}}{2} \left( \frac{1}{\Delta_{2}}-\frac{1}{\Delta_{1}}\right)$ and $\Omega_{e2}=\frac {\Omega_{2}^{*}\Omega_{1}}{2}\frac{1}{\Delta_{2}}$, respectively. The projection operators are also rewritten using the on-site hard-core bosonic operators: $\sigma_{z}=\left ( | e \rangle \langle e | - | g \rangle \langle g | \right )$, $\sigma_{(+)} = | e \rangle \langle g |$, and $\sigma_{(-)} = | g \rangle \langle e|$. Additionally, $n^{i} = \frac{1 + \sigma_{i}^{z}}{2}$, $\omega_{A} = \omega_{1} + \Delta_{2}$. 

We turn to a discussion of the properties of the effective Hamiltonian.
It is comprehensive to rewrite the full effective Hamiltonian,
\begin{align}
     H_{eff} =&\omega_{c}a^{\dagger}a + (\omega_{A}+A\cos{\omega_{s}t}) \sum_{j=1}^{N} n_{j} + V\sum_{\langle i,j \rangle} n_{i}n_{j}\nonumber\\ 
     +&\sum_{j=1}^{N} \left (\frac{\Omega_{e1}}{\sqrt{N}} a\sigma^{(+)}_{j}+\frac{\Omega_{e2}}{\sqrt{N}} a^{\dagger} \sigma^{(+)}_{j} + h.c.\right ) \label{fHamil}
\end{align}
Here we are facing a similar time-dependent periodic issue as that in Eq.\eqref{EDri1}.
Following the same procedure, we first apply the unitary transformation
$ U_{1} = \exp \left \{-i\omega_{C}a^{\dagger}a t-i \left(\omega_{A}t + \frac{A}{\omega_s} \sin{\omega_{s} t} \right) \sum_{j=1}^{N} n_{j}\right \}$,  under which the Hamiltonian becomes 
\begin{align}
    \widetilde{H}_{E} =& V\sum_{\langle i,j \rangle}n_{i}n_{j}
    + \frac{\Omega_{e1}}{\sqrt{N}} \sum_{j=1}^{N} \left \{ W_{1}(t)a\sigma_{j}^{+} + h.c. \right \}\nonumber\\
    +& \frac{\Omega_{e2}}{\sqrt{N}} \sum_{j=1}^{N} \left \{ W_{2}(t)a^{\dagger}\sigma_{j}^{+} + h.c. \right \}\label{Hee}
\end{align} 
Here, $W_{1}(t) = \exp \left [i \left (-\omega_{C}t + \omega_{A}t + \frac{A}{\omega_s}\sin\omega_{s}t \right ) \right ]$ and $W_{2}(t) = \exp \left [i\left (\omega_{C}t + \omega_{A}t + \frac{A}{\omega_s}\sin\omega_{s}t\right )\right ]$. 
In the condition of $ \left |\omega_{A} - n_t \omega_s \right | \ll \omega_s$, 
we take the resolved side-band approximation, which means that only the $ n_t$th side-band needs to be considered. 
Based on this consideration, we next transform the Hamiltonian Eq.\eqref{Hee} back by $ U_{2} = \exp \left[ i \left( \omega_{C} - n_{t}\omega_{s} \right) a^{\dagger}at + i\left( \omega_{A} - n_{t}\omega_s \right) t \sum_{j=1}^{N}n_{j} \right ]$, which leads to
\begin{align}
    &H_{F} =\frac{\Omega_{e1}}{\sqrt{N}} \sum_{j=1}^{N} \Big\{\exp \left( i\frac{A}{\omega_s}\sin\omega_{s}t \right) a\sigma_{j}^{+} + h.c. \Big\}\nonumber\\
    +&\frac{\Omega_{e2}}{\sqrt{N}} \sum_{j=1}^{N} \Big\{\exp \left[i \left(\frac{U} {\omega_s} \sin\omega_{s}t + 2n_{t}\omega_s t \right) \right] a^{\dagger}\sigma_{j}^{+} + h.c. \Big\}\nonumber\\
    &+ \left( \omega_{C} - n_{t} \omega_{s}\right) a^{\dagger}a + V\sum_{\langle i,j \rangle}n_{i}n_{j} + (\omega_{A} - n_{t}\omega_{s})\sum_{j=1}^{N}n_{j} 
\end{align}  

Using the Jacobi-Anger relation $ e^{iK \sin\omega_s t} = \sum_{n=0}^{\infty}J_{n}[K]e^{i n \omega_s t}$ and the Floquet Magnux expansion, we finally obtain the lowest order effective Hamiltonian,
\begin{align}
	H_{FM} =& \frac{1}{T} \int_{0}^{T} H_{F} dt =  \Omega_{e3}\sum_{j=1}^{N} \sigma_{j}^{+} a 
	+ \Omega_{e4} \sum_{j=1}^{N}\sigma_{j}^{+} a^{\dagger} \nonumber\\
    +& \widetilde{\omega}_{C} a^{\dagger}a + V\sum_{\langle i,j \rangle }n_{i}n_{j} + \widetilde{\omega}_{A}\sum_{j=1}^{N} n_{j}+ h.c. \label{HFM}
\end{align}  
where 
\begin{align}
    \Omega_{e3} =& \frac{\Omega_{e1}}{\sqrt{N}} J_{0}[\frac{A}{\omega_s}], 
    \Omega_{e4} = \frac{\Omega_{e2}}{\sqrt{N}}J_{2n_t}[\frac{A}{\omega_s}]\\
    \widetilde{\omega}_{C} =& \omega_{C} - n_{t}\omega_{s}, 
    \widetilde{\omega}_{A} = \omega_{A} - n_{t} \omega_{s}
\end{align}

\begin{figure}[t]
    \centering
    \includegraphics[width=1.04\linewidth]{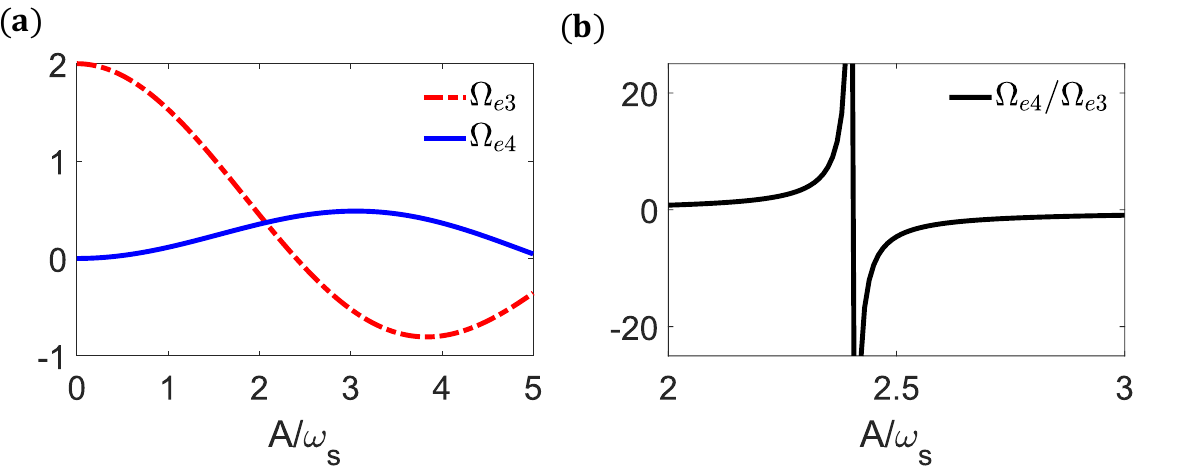}
    \label{effec_coupl}
    \caption{(a) The coupling strength of RW  and  CRW versus the rate of driving amplitudes to frequencies, denoted by blue solid line and red dotted solid line, respectively. 
    (b) The ratio of the coupling strength of CRW to RW against the one of driving amplitudes to frequencies. Here we set all the parameters in units of $\Omega_{e2}$, $\Omega_{e1}=2$. The order of the Bessel function $n_{t} = 1$ and the total particle size $N=1$.}
    \label{coupling}
\end{figure}

Note that $\Omega_{e1}$ and $\Omega_{e2}$ in Eq.(18) depend on the detunings of the two-photon process, which must meet some approximation conditions, it is difficult to regulate $\Omega_{e3}$ and $\Omega_{e4}$ through them.
However, by introducing the Bessel function, our Floquet engineering scheme makes the controllability of the effective coupling strength of RW and CRW possible.
The numerical results for demonstrating their tunability are shown in Fig.\ref{coupling}, where $\Omega_{e1} = 2$, $ \Omega_{e2}=1$ and $ n_{t} = 1$.
For convenience of following discussions, we denote $\Omega$ as the mean coupling strength of atoms and the cavity field, $\alpha=\frac{2\Omega_{e4}}{\Omega_{e3}+\Omega_{e4}}$. Then Eq.\eqref{HFM} is rewritten as
\begin{align}
	H =& \widetilde{\omega}_{c} a^{\dagger}a + (1-\alpha) \frac{\Omega}{\sqrt{N}} (\sum_{j=1}^{N} \sigma_{j}^{+} a + a^{\dagger}\sigma_{j}^{-})\nonumber\\
    +& \alpha \frac{\Omega}{\sqrt{N}} (\sum_{j=1}^{N}\sigma_{j}^{+} a^{\dagger} +a\sigma_{j}^{-}) + V \sum_{\langle i,j \rangle}n_{i}n_{j} + \widetilde{\omega}_{A}\sum_{j=1}^{N}n_{j} \label{E8}
\end{align}
For $V = 0$, Eq.\eqref{E8} reduces to the ADM where the critial point of the phase transition from the normal phase to the SR phase, $\Omega = \sqrt{\widetilde{\omega}_{C}\widetilde{\omega}_{A}}$, is independent of $\alpha$\cite{adm2, Daspra}.
When Rydberg dipole-dipole interaction is incorporated but with $\alpha=0$, previous research has demonstrated the quantum phase transition in both one dimension and square arrays, and even discovered a novel phase: the SRS phase which breaks $U(1)$ and translation symmetries simultaneously\cite{srs1,srs2}. 
It is equivalent to the Ising-Dicke model for $\alpha = 0.5$.
In addition to observation of quantum phase transitions\cite{idqpt1, idqpt2, isingm}, meson oscillations in the spins and squeezed-vacuum light state\cite{cav7}, some results in such a case can find applications in enhancing the metrology sensitivity of temperature and external fields\cite{Gammelmark}. 

\section{Adiabatic SR and SRS State Preparation}
\label{Adiabatic}
Having discussed engineering ADM, we now demonstrate that the tunable ratio of CRW to CR can facilitate the adiabatic state preparation. 
One method for preparing the ground state of complex systems is to adiabatically evolve from an easily accessible ground state.
For Rydberg atom systems in optical cavities, the normal state can be chosen as the initial ground one.
We focus on the target ground states of SR and SRS ones, which both exist in the present ADM with the Rydberg atomic interaction.

Before delving into the impact of CRW on the state preparation, we note that the Hamiltonian Eq.\ref{E8} remains invariant under the transformation $(a, a^{\dagger}) \to (-a, -a^{\dagger})$ and $(\sigma^{-}_{j}, \sigma^{+}_{j}) \to (-\sigma^{-}_{j}, -\sigma^{+}_{j})$.
It provides insights that there is one conserved quantity, i.e. the parity, which is constructed as $P = (-1)^{a^{\dagger}a}\prod_{j=1}^N\sigma^{z}_{j}$.  
Assuming the time-driven Hamiltonian keeps the parity symmetry in the whole evolution process, then one instantaneous eigenstate will always maintain a fixed parity.
According to recent research of symmetry-protected quantum adiabatic evolution\cite{arQDE}, the population transfer between different eigenstates with different parity is exactly forbidden. 
Only the state transition of the same parity is allowed. 
The following discussion will exploit this principle.

We employ the Runge-Kutta method and exact diagonalization (ED) to simulate the adiabatic evolution. 
The Runge-Kutta method is utilized to evolve the wave function based on the instantaneous Hamiltonian, while exact diagonalization is employed to obtain the eigenstates of the instantaneous Hamiltonian.
For simplicity, the numerical simulation is conducted on the one-dimensional chain, and the results are similarly applicable to the square lattice.
Due to the exponential growth of the Hilbert space with the number of particles, we simulate the system with atoms $N=6$ and photon cutoff numbers $N_{a}=40$.
During the process, we monitor various quantities: the fidelity $F=|\langle \psi(t) | \psi_e(t) \rangle|^{2}$ between the propagated state $\psi(t)$ and the instantaneous eigenstates $\psi_e(t)$, the mean number of cavity photons$N_{a} = a^{\dagger}a$ and the structure factor $S(q)/N =\langle |\sum_{j=1}^{N} P_{j} e^{iqr_{k}}|^{2} \rangle /N^{2} $ with $q = \pi$.
Here $N_{a}$ and $S(q)$ help us identify the characteristics of eigenstates we care about.

\begin{figure}[!t]
    \centering
    \includegraphics[width=1.0\linewidth]{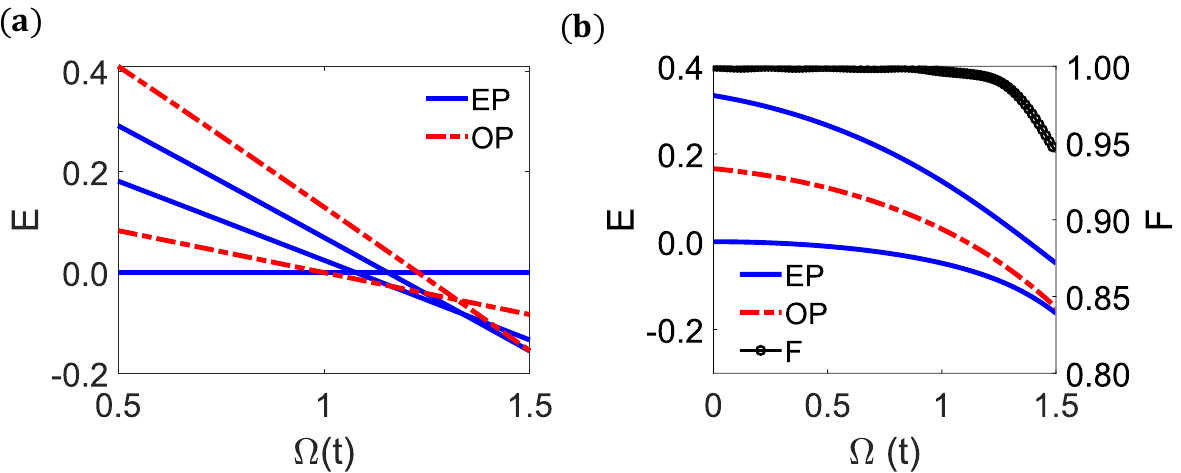}
    \caption{(a) Some low energy spectrum as a function of coupling strength $\Omega(t)$ for $\alpha=0$.
    (b) Low energy spectrum and fidelity(black circle line, denoted by F) versus $\Omega(t)$ for $\alpha=0.7$.
    Here, the blue solid line and the red dotted solid line represent even parity (EP) and odd parity (OP), respectively, corresponding to the eigenvalues of the $P$ operator being $1$ and $-1$.
    Unit with Rydberg interaction strength $V$, we set $\widetilde{\omega}_{c}=1$ and $\widetilde{\omega}_{A}=1$.
    We retain the Hilbert space with the photon number $N_{a}\leq 40$, the lattice size $N=6$.}
    \label{SR}
\end{figure}

Fig.\ref{SR} shows some instantaneous eigenenergies related to the SR state preparation, where a sweeping protocol is chosen as $\Omega(t)=\Omega t /T_{SR}$ with $T_{SR}$ the required preparation time. 
Experimentally, such a process can be achieved by gradually increasing the intensity of classical light.
Due to the parity symmetry, the instantaneous eigenstates will possess even or odd parity, see the corresponding blue solid line or the red dashed line in Fig.\ref{SR}.
When $\alpha=0$, which implies vanished CRW, one observes some energy level crossings of even-parity eigenstates from Fig.\ref{SR}(a).
Since the adiabatic evolution condition is violated in such cases, the target ground state with even parity can never be well adiabatically prepared from the initial even-parity ground state.
With the increase of $\alpha$, however, some gaps will develop between those even-parity energy spectrums,  as shown in Fig.\ref{SR}(b).
Although the quasi-degeneracy is evidenced at the final moment $\Omega(t)=1.5$, the odd-parity instantaneous eigenstate will never be occupied if one chooses the lowest-energy even-parity adiabatic passage.
In Fig.\ref{SR}(b), we also present the calculated fidelity for the entire instantaneous process, which can clearly corroborate our discussion.
Therefore, by tuning $\alpha$, the adiabatic evolution condition can be satisfied and the SR state preparation can be completed.

\begin{figure}[H]
    \centering
    \includegraphics[width=1.0\linewidth]{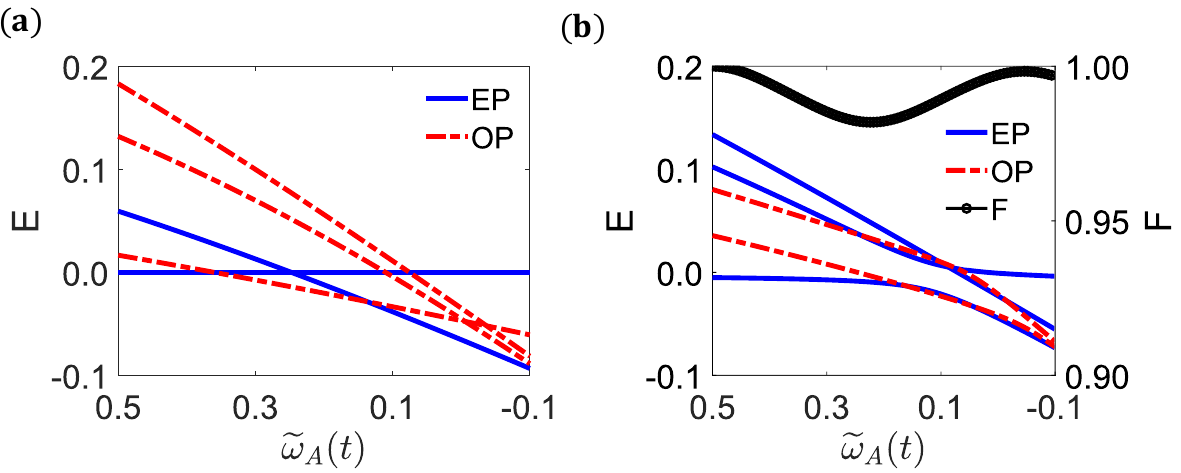}
    \caption{(a) The function of low energy spectrum to  $\widetilde{\omega}_{A}(t)$ for $\alpha=0$.
    (b) Dependence of low energy spectrum on $\widetilde{\omega}_{A}(t)$ with $\alpha=0.3$.
    Right $y$ axis of (b) is the fidelity (F, black circle line), which is the overlap of propagated state and the first even parity instantaneous eigenstates.
    Even parity(EP) and odd parity(OP) are labeled by blue solid line and red dotted solid line, respectively.
    For other parameters, we set $V=1$, $\Omega = 0.6$, and $\widetilde{\omega}_{C} = 1$.}
    \label{SRS}
\end{figure}

The discussion of the SRS state preparation can be proceeded similarly. 
Upon linearly sweeping $\widetilde{\omega}_A$ from 0.5 to -0.1, some instantaneous low-energy spectrum is shown in Fig.\ref{SRS}.
We first focus on the limitation $\alpha=0$.
Given the initial groud state and the target one share the same even parity, the even-parity energy level crossing shown in Fig.\ref{SRS}(a) indicate that it is impossible to transfer to the SRS state with high fidelity through the adiabatic evolution.
According to the principle of symmetry-protected transition again, we now disregard the odd-parity eigenstates because of their different parity from the target one.
Just as shown in Fig.\ref{SRS}(b), the level crossing can be avoided with opening a finite gap between the instantaneous eigenstates of the even parity when increasing $\alpha$.
Then the adiabatic evolution may always appear if our system is driven sufficiently slowly.
Our numerical fidelity in Fig.\ref{SRS}(b) has demonstrated most of population staying in the instantaneous lowest-energy eigenstate of the even parity, which indeed confirms our analysis.

In summary, we can regulate the distribution of instantaneous eigenenergies by tuning the ratio of CRW to CR, $\alpha$, thus facilitating the preparation of the state SR and SRS.
In particular, by employing the principle of symmetry-protected transitions, our tunning strategy aims at pushing away the eigenstates with the same parity as the target adiabatic pathway, while ensuring the eigenstates closed to the pathway having different parity. 
In this way, roughly speaking, the efficiency of adiabatic state preparation is determined by the energy gap between nearest neighboring instantaneous eigenstates of the same parity.

\section{Conclusion and Outlook}
\label{outlook}
We have designed and Floquet engineered the anisotropic Dicke model based on the Rydberg arrays in cavity.
Also, we have demonstrated the possibility to manipulate the ratio of the coupling strengths of the RW terms to the CRW ones by appropriately adjusting the driving pumping laser.
As a showcase, our adiabatic evolution results have elucidated the superior control strategy for synthesizing complex quantum states from simpler ones.
This level of control is not only crucial to tailoring the behavior of the system in a targeted manner, but also has great potential for applications even beyond the realm of physics, for instance, combinatorial optimization problems\cite{QOW, bomb}.

\section{ACKNOWLEDGMENTS}
This work was supported by the National Science Foundation of China under Grant No. 12274045. T.W. acknowledges funding from the National Science Foundation of China under Grants No. 12347101, Chongqing Natural Science Foundation under Grants No. CSTB2022NSCQ-JQX0018, and the Program of State Key Laboratory of Quantum Optics and Quantum Optics Devices under Grant No. KF202211.

\bibliographystyle{apsrev4-1}
\bibliography{referen}

\end{document}